\documentclass[12pt,preprint]{aastex}
\usepackage{emulateapj5}
\usepackage{epsf}
\usepackage{flushrt}
\usepackage{amsfonts}

\newcommand{\pow}[2]{\ensuremath{\mbox{#1}^{\rm #2}}}

\newcommand{\water}{H$_2$O}
\newcommand{\kmS}{km \pow{s}{-1}}

\submitted{Submitted March 6, 2002; Accepted for publication in 
ApJ(Letters)}

\shortauthors{Watson, Sarma, \& Singleton}
\shorttitle{Gaussian Spectral Line Profiles of Astrophysical Masers}

\begin{document}

\title{Gaussian Spectral Line Profiles of Astrophysical Masers} 

\author{W.\ D.\ Watson\altaffilmark{1}, A.\ P.\ Sarma\altaffilmark{2}, 
M.\ S. Singleton\altaffilmark{1}}

\altaffiltext{1}{Physics Department, University of Illinois at
Urbana-Champaign, 1110 W. Green St., Urbana IL 61801;
w-watson@uiuc.edu}

\altaffiltext{2}{NCSA/Astronomy Department, University of Illinois at
Urbana-Champaign, 1002 W. Green St., Urbana IL 61801; 
asarma@astro.uiuc.edu}

\begin{abstract}

Calculations are performed to demonstrate the deviations from Gaussian
that occur in the spectral line profiles of a linear maser as
a result of the amplification process. Near-Gaussian profiles are
presented for bright, interstellar 22 GHz water masers obtained from
high resolution Very Long Baseline Array (VLBA) observations of W3 IRS 5.
For the profiles to be so close to Gaussian, the calculations indicate
that these masers must originate in quite hot gas with temperatures
greater than 1200 K --- a conclusion that is supportive of C-type shocks
as the origin of these masers. In addition, the degree of saturation of
these masers must be less than approximately one-third, from which it
follows that the beaming angles are less than about
\pow{10}{-4}\ ster and the actual luminosities are modest. If spectral
profiles that are as close to Gaussian as the profiles presented in this
initial investigation are found to occur widely, they can be valuable
diagnostics for the environments of astrophysical masers.

\end{abstract}


\section{INTRODUCTION}

With increasing intensity, the spectral line profiles that are calculated
for astrophysical masers first become narrower and then rebroaden when
the masers become saturated. At high saturation, the profile in Doppler
velocity becomes the same as the Maxwellian (Gaussian) velocity
distribution of the masing molecules. This behavior has been established
to be independent of whether the masers are treated as linear 
(\citealt{lit71}), or as having finite cross sections such as disks
and spheres (\citealt{ew94}; Wyld \& Watson 2002, in preparation). It
seems to be less widely recognized that the profiles also can be
negligibly different from Gaussians at low intensities where the maser
is unsaturated. We demonstrate how the deviations of the profile from 
Gaussian can (along with the breadth of the profile) then be used to
obtain information about the masers. In particular, lack of knowledge
about the solid angle into which the maser radiation is beamed
--- which is necessary to obtain the luminosity and the energy density
from the observed radiative flux --- has been an outstanding weakness in
the interpretation of astrophysical masers.

At low angular resolution, the observed spectral line profiles are not
expected to be indicative of the profiles of individual masers because
multiple masing components may be present within a telescope beam. However,
the profiles presented here have been observed at very high angular
resolution with the Very Long Baseline Array (VLBA) and are quite close to
Gaussian. Most likely, the deviations from Gaussian in these profiles are
upper bounds to the actual deviations in the profiles of individual masers.

\section{OBSERVATIONS OF NEAR-GAUSSIAN MASER PROFILES}
\label{sOBS}

\begin{figure*}
\epsscale{1.5}
\plotone{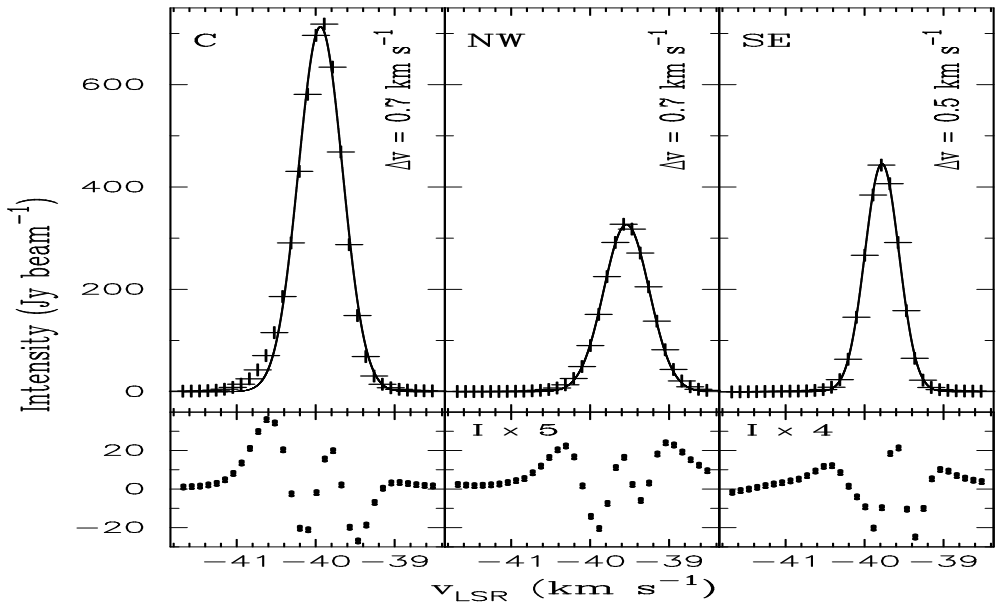}
\caption{VLBA 22 GHz \water\ maser profiles toward W3 IRS 5. The
crosses represent the observed data, and the smooth curves represent the
fitted Gaussians. The dots in the lower panel represent the residuals
from the Gaussian fit that have been estimated at each observed velocity.
The error bar along the intensity axis at each observed point (as measured
from the rms in the images of Stokes V = right minus left circular 
polarization) is 16 mJy \pow{beam}{-1}. The linewidths (FWHM) are also
given in the figure. Other details of the observations are given 
in \citet{str01}. The maser labels used are also taken from that paper.
Note that the residuals for the masers NW and SE have been multiplied
by factors of 5 and 4 respectively to increase their visibility.
\label{fMP}}
\end{figure*}

The 22 GHz \water\ maser profiles toward W3 IRS 5 shown in Figure\
\ref{fMP}\ were obtained with the VLBA in order to measure the circular
polarization of the maser radiation, and hence to infer the strengths of
the magnetic fields. Also shown in this Figure are the Gaussian fits and
the differences between the fits and the data. The observational procedures 
to obtain these profiles have been described in a previous Letter
(\citealt*{str01}). Here we examine the near-Gaussian nature of the
profiles. W3 IRS 5 is an infrared double source located between the 
\ion{H}{2}\ regions A and B in the W3 giant molecular cloud 
(e.g., \citealt{tgc97}). \citet{cgjw94}\ found seven hypercompact
continuum sources (diameter $<$ 700 AU: \citealt{tgc97}) in the W3 IRS 5
region with the Very Large Array (VLA). They interpreted these sources as
stellar winds from embedded B stars (cf.\ \citealt{tgc97}). These authors
also found that some of the brightest \water\ masers coincide with the
locations of 15 GHz continuum images of the hypercompact sources. From the
flux in the observed profiles and the maser sizes (assumed unresolved;
hence sizes are the half-power beamwidth, which is $\sim$0.7 mas), we find
that the minimum brightness temperatures $T_B$ for the masers in Figure
\ref{fMP}\ are 1 to 4 $\times$ \pow{10}{12}\ K.

\section{CALCULATIONS}

The equation of transfer for the dimensionless intensity $I(v)$ of a
ray of maser radiation at Doppler velocity $v$ can be expressed as a
function of the optical depth parameter $s$ in the linear maser
idealization as

\begin{equation}
\frac{dI}{ds}=\left[  \frac{I}{1+\left(  1+g_u/g_l\right)
I}+S\right]  \exp\left(  -v^{2}/0.36v_{t}^{2}\right)
\label{e1}
\end{equation}

\begin{figure*}
\begin{minipage}[t]{0.48\textwidth}
\begin{center}
\epsscale{0.6}
\plotone{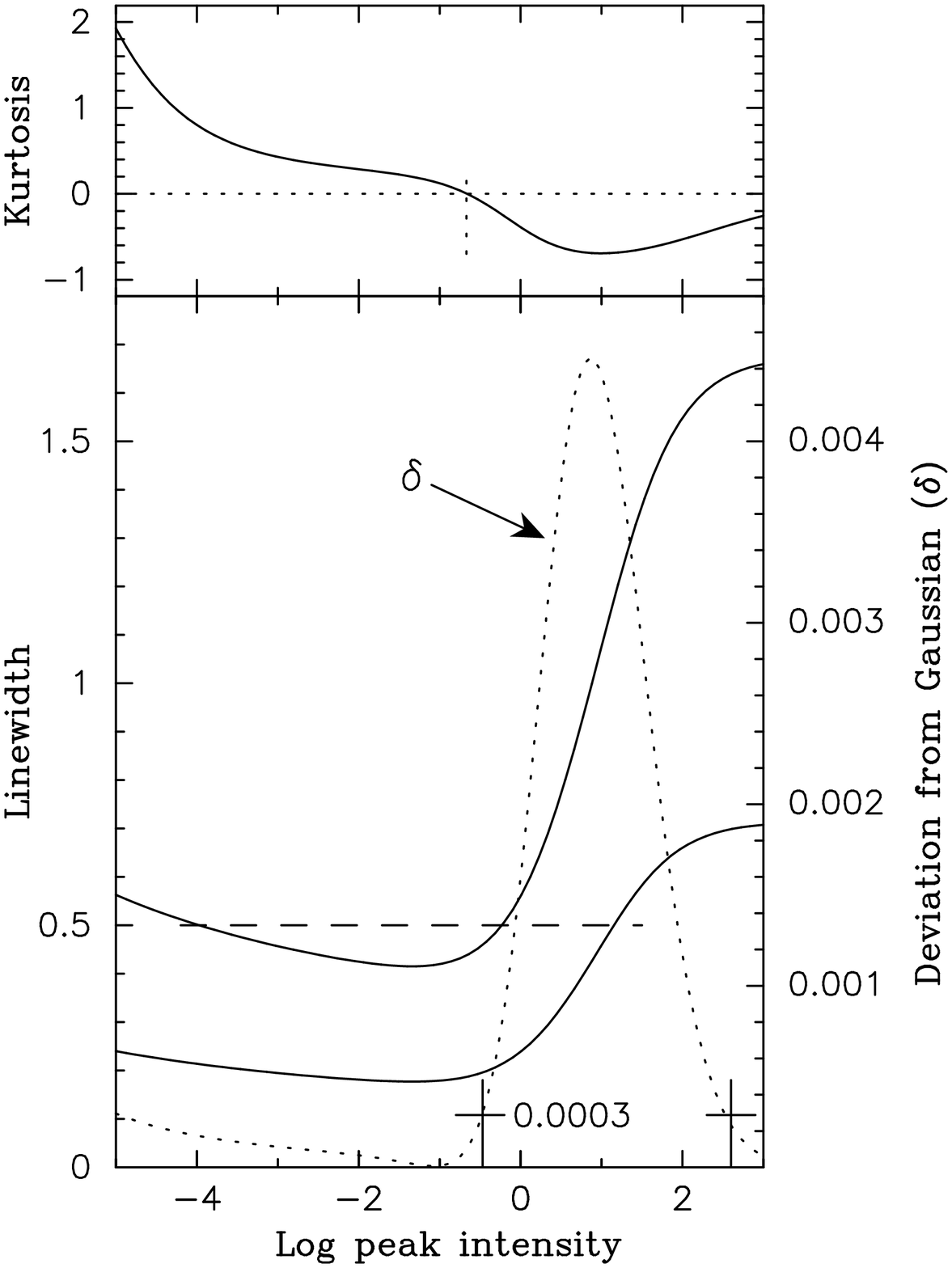}
\caption{The lower panel shows the representative linewidths
$\Delta v_{1/2}$ (arbitrary units) that are calculated as a function of
the log of the peak intensity $I_p$ for masing gas at two gas temperatures.
The horizontal line at $\Delta v_{1/2}=0.5$ represents an observed
linewidth. Its intersections with the curves for the calculated
linewidths determine the gas temperatures and intensities that are
compatible with this linewidth. The deviation of the profile from a
Gaussian as measured by $\delta$ is given by the dotted curve on which
the crosses mark $\delta$ = 3 $\times$ \pow{10}{-4}\ that is representative
for the observed profiles in Figure\ \ref{fMP}. Though the results of
calculations are presented in this figure only for $I_c$ = \pow{10}{-9},
they are indicative for the range of choices for $I_p$ that are relevant
here (i.e., those in Figure\ \ref{fIvsT}). The upper panel shows the
kurtosis for these profiles. Note that $\delta$ and the kurtosis do not
depend on the temperature of the gas.
\label{fKL}}
\end{center}
\end{minipage}\hfil
\begin{minipage}[t]{0.1\textwidth}
\end{minipage}
\begin{minipage}[t]{0.48\textwidth}
\centering
\epsscale{0.55}
\plotone{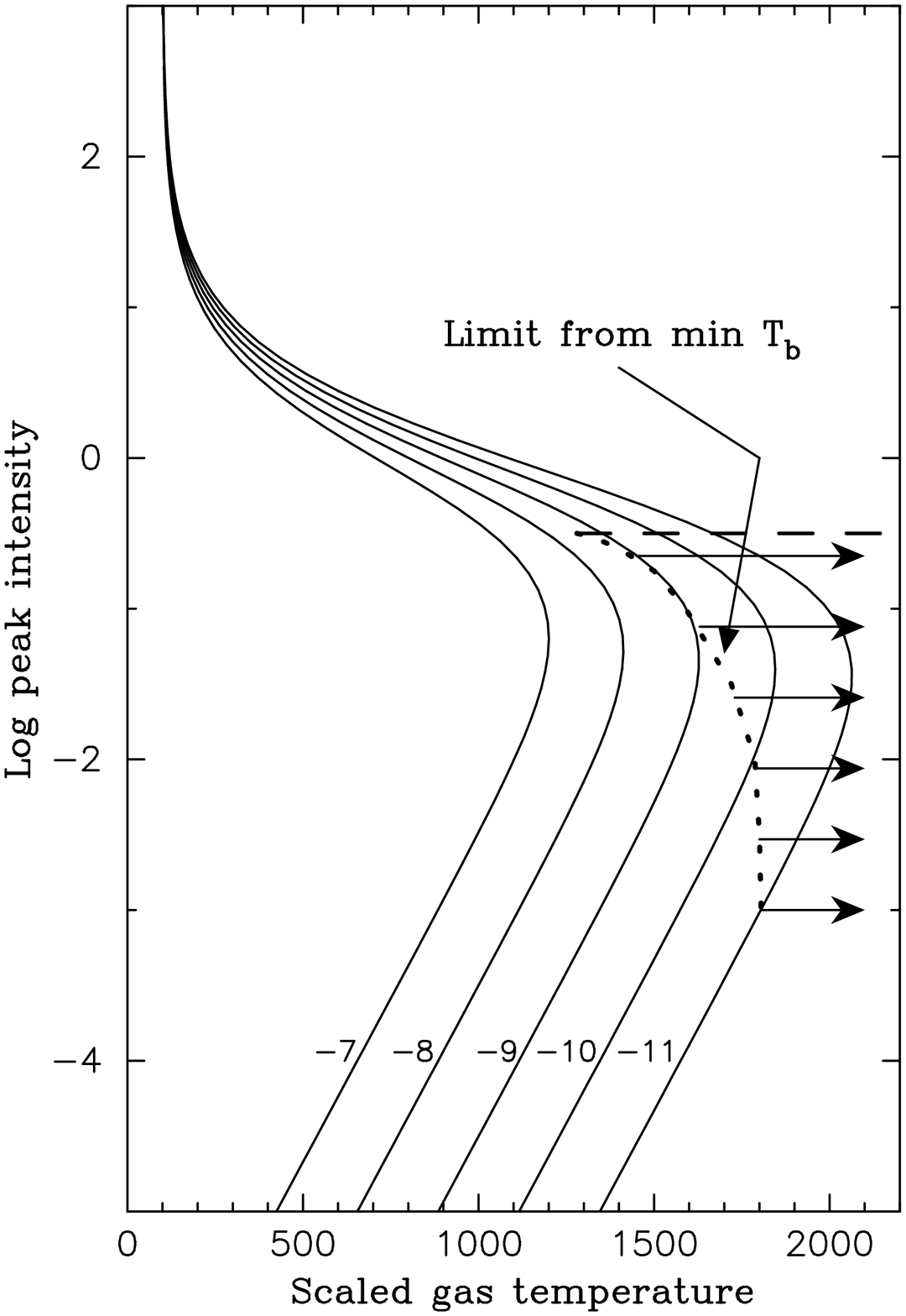}
\caption{The scaled gas temperature and the log of the peak intensity
$I_p$ which yield a spectral linewidth $\Delta v_{1/2}$ (FWHM) in
calculations analogous to those for Figure\ \ref{fKL}. The scaled gas
temperatures must be multiplied by the factor $t$ =($\Delta v_{1/2}$/0.5
\kmS)$^{2}\ \times$(M/M$_{{\rm H}_2{\rm O}}$) to convert them to actual
gas temperatures in Kelvins, where M is the mass of a molecule of the
masing species. The various curves (solid lines) are computed for
different choices of $I_c$ = \pow{10}{-7}, \pow{10}{-8}, \pow{10}{-9},
\pow{10}{-10}, \pow{10}{-11}\ as indicated. The horizontal dashed line
at log $I_p$ = $-$0.5 marks the approximate upper limit in Figure\ 
\ref{fKL}\ for the near-Gaussian profiles for which 
$\delta$ $\lesssim$ 3 $\times$ \pow{10}{-4}. The dotted line represents
$I_c$ = \pow{10}{-8}$I_p$, which follows from the lower limit to the
observed brightness temperatures for the masers in \S\ \ref{sOBS}.
According to the interpretation, the allowed gas temperatures must be on
the high temperature side of this line as indicated by the arrows.
\label{fIvsT}}
\end{minipage}
\end{figure*}

Here, the $I$ =\ $I(v)$ is in units of the saturation intensity 
$I_s=(8\pi h\nu^{3}\Gamma/c^{3}g_{u}A\Delta\Omega)$, where $\nu$ is
the frequency of the transition, $\Gamma$ is the ``phenomenological''
decay rate of the molecular state, $g_u, g_l$ are the degeneracies of the
upper and lower states, $A$ is the Einstein A-value, and $\Delta\Omega$
is the beaming angle for the maser radiation. Also, the source function
due to spontaneous emission
$S=[(g_{u}A\Delta\Omega/4\pi\Gamma)(\Lambda/\Delta\Lambda)]$ where
$\Lambda$ is the phenomenological pumping rate into the upper state and
$\Delta\Lambda$ is the difference between the pumping rates into the
upper and lower states. The dispersion (FWHM) in the thermal velocities
of the masing molecules is $v_{t}$. If spontaneous emission is ignored,
equation\ \ref{e1}\ can be integrated to yield 
$I\exp[(1+g_u/g_l)I]=I_{c}\exp[s\exp(-v^{2}/0.36v_{t}^{2})]$ for
$I_{c} \ll I$, where $I_{c}$ is the intensity of the continuum radiation
that is incident at the far end of the maser which serves as the ``seed''
radiation for the maser. The $I(v)$ that is found by integrating equation
\ref{e1}\ is fit to a Gaussian using the least squares procedure. The
result $\delta$ of dividing the integral of the squares of the normalized
deviations by the linewidth $\Delta v_{1/2}$ (FWHM) of the computed profile
\begin{equation}
\delta=\{\int[I(v)-a_{1}\exp(-v^{2}/a_{2})]^{2}dv\}/I_{p}^{2}\Delta
v_{1/2}
\label{e2}
\end{equation}
is then adopted as a quantitative measure of the deviation of the spectral
line profile from a Gaussian. Here, $a_{1}$ and $a_{2}$ are the parameters
of the Gaussian that are obtained in the least squares fitting and
$I_{p}$ $\equiv I(v=0)$ is the peak intensity within the spectral line.
Note that $\delta$ is independent of $v_{t}$ (that is, of the kinetic
temperature of the gas) and that $\Delta v_{1/2}$ $=$ $v_{t}$ $\times$ 
$f$($I_{c},I_{p})$ when $S$ is ignored. The function $f$($I_{c},I_{p})$
$\longrightarrow1$ as $I_{p}$ (which also is a measure of the degree of
saturation) becomes much greater than one. In Figure\ \ref{fKL}\, examples
of the variations of the calculated $\delta$ and $\Delta v_{1/2}$ are
shown as a function of the peak intensity $I_{p}$ (or the degree of
saturation) for $I_{c}$ = \pow{10}{-9}, which we will reason subsequently
is representative for the data in \S\ \ref{sOBS}. Clearly, the main
deviations in $I(v)$ from a Gaussian occur in the region of partial
saturation. The two curves for the linewidth have different values for
$v_{t}$, and are thus shifted vertically from one another by a constant
factor. The horizontal line in Figure\ \ref{fKL}\ intersects these two
curves at different points, and is used to indicate that a specific
linewidth ordinarily is compatible with a range of gas temperatures and
a range of degrees of saturation. The intersections determine the
corresponding $I_{p}$ and $\delta$ when a linewidth of 0.5 results from
gas at the two temperatures. In Figure\ \ref{fIvsT}, the relationship is
shown between the intensity $I_{p}$ and the gas temperature (scaled)
which yield a specific linewidth $\Delta v_{1/2}$, for several values
of $I_{c}$. Note that the information in Figure\ \ref{fIvsT}\ can be
scaled as described in the caption for any linewidth and molecular
species. The ``kurtosis'' --- a quantitative measure of whether the
profile is more [positive] or less [negative] peaked than the Gaussian
at line center (e.g., \citealt*{pft86}) --- also is shown in 
Figure\ \ref{fKL}. That the kurtosis changes sign at essentially the
value of $I_{p}$ where the linewidth is a minimum and where the
saturation begins to be significant is a further indication of the
change in the character of the spectral line at this point.

We find that, at least for the ranges that are considered here for the
parameters, spontaneous emission yields results that differ negligibly
from those obtained for incident continuum radiation with $I_{c} =S$. We
thus present results labelled only according to the $I_{c}$, but with
the understanding that they are applicable for $S = I_{c}$ as well.

\section{INTERPRETATION OF THE 22 GHz \water\ MASER PROFILES}

The relevant characteristics for the profiles in Figure\ \ref{fMP}\ are
($\Delta v_{1/2}$ \kmS, $\delta$) = C\ (0.7, 2.4 $\times$ \pow{10}{-3}),
NW (0.7, 2.4 $\times$ \pow{10}{-4}), SE (0.5, 1.8 $\times$ \pow{10}{-4}), 
and for a fourth profile observed by \citet{str01}\ but not shown in the
figure, (0.5, 8.5 $\times$ \pow{10}{-4}). At least within the relevant
range of values for $I_{c}$ that is considered in Figure\ \ref{fIvsT},
the values of $I_{p}$ at which specific $\delta$ occur are essentially
independent of $I_{c}$ when $\delta \lesssim$ \pow{10}{-3}. Hence, to
be compatible with the observed profiles in \S\ \ref{sOBS}\ for which
three have $\delta$ $\lesssim$ \pow{10}{-3}, the intensity must either
be quite large ($\gtrsim$ 100) or it must be somewhat less 
than 1 according to Figure\ \ref{fKL}. We reject the high saturation
solutions because (i) the required saturation seems excessive, but more
objectively because (ii) the gas temperatures $T$ that would be 
indicated ($\simeq$ 100 K for $\Delta v_{1/2}$ = 0.5 \kmS) in 
Figure\ \ref{fIvsT}\ are below what seems to be required for the
chemistry and pumping of these masers, and (iii) the 22 GHz \water\ 
maser feature probably consists of a blend of three hyperfine
components. Hyperfine structure is omitted here, but would cause
irregular profiles at temperatures of a few hundred K. At these
temperatures, $v_t$ is too small in comparison with the hyperfine
splittings for the components to merge completely. In addition, (iv) the
``kurtosis'' of the four line profiles that are observed is positive as 
is expected from Figure\ \ref{fKL}\ for maser profiles at 
$I_{p}\ \lesssim$ 0.3, but not for larger $I_{p}$. Also, the residuals
at the intensity peak for two of the three profiles in Figure\ \ref{fMP}\
(and for a fourth profile that is not shown) are positive, directly
indicating that the observed profiles are more sharply peaked (positive
kurtosis) than the Gaussian fit. For the third profile shown in Figure\
\ref{fMP}\ (maser SE), the peak in the residual is shifted by one velocity
channel from the peak in the observed profile.

Now consider the likely ranges for $I_{c}\ \simeq$ 
\pow{10}{-9}$T_c\Delta\Omega$ and $S \simeq$ 
\pow{10}{-9}$\Delta\Omega(\Lambda/\Delta\Lambda)$,
where the background continuum intensity has been expressed as a
brightness temperature $T_c$ and the usual estimate for the 22 GHz masers
$\Gamma$ = 2 \pow{s}{-1}\ has been adopted. In calculations for the pumping
of these masers, the fractional inversion $\Delta\Lambda/\Lambda$ typically
is a few percent (e.g., \citealt{aw93}) to give 
$S\ \simeq$ \pow{10}{-7}$\Delta\Omega$. Plausible values for $T_c$ range
from little more than the 3K cosmic background to the \pow{10}{4}\ K that
is characteristic of \ion{H}{2}\ regions. Then, since it is unimportant
for the calculated spectra whether the seed radiation is provided by
spontaneous or by background radiation, for convenience we can express
the likely range for the seed radiation in terms of $I_{c}$ alone with
\pow{10}{4}\ $\gtrsim$ $T_c$ (K) $\gtrsim$ \pow{10}{2}. Because $I_p$ can
be related to $\delta$, it is useful to eliminate $\Delta\Omega$ and
express $I_c$ as $I_{p}T_{c}/T_{B}$ = 
$10^{-8}I_{p}[(T_{c}/10^{4}$ K)/($T_{B}/10^{12}$ K)], or
$I_{c} \leq 10^{-8}I_{p}$ here, since $T_{B}$ = \pow{10}{12}\ K is a
lower limit for the data in \S\ \ref{sOBS}. The locations on
the curves in Figure\ \ref{fIvsT}\ where $I_{c} = 10^{-8}I_{p}$ are
indicated by the dotted line. The upper end of this line is at 
$I_{p}$ = 0.3 (and $I_{c}$ = 3 $\times$ \pow{10}{-9}) --- the value at
which $\delta$ = 3 $\times$ \pow{10}{-4}\ in Figure\ \ref{fMP}.
Hence, the allowed region in Figure\ \ref{fIvsT}\ is that indicated by
the arrows, and this dotted line indicates the lowest gas temperatures
that are compatible with the two profiles in \S\ \ref{sOBS}\ that are
closest to Gaussian profiles --- and probably with the other two profiles
as well in view of the sense of the uncertainties in $T_c$, $T_B$ and
$\delta$. In turn, $\Delta\Omega$ can be related to the intensities and
brightness temperatures of the masers through $\Delta\Omega$ = 
(4$\pi h\nu\Gamma/ckg_{u}A)I_{p}/T_{B} \simeq 3 \times 
10^{-4}I_{p}/(T_{B}/10^{12}$ K) ster. 

It is evident that the foregoing analysis indicates that the maser profiles
described in \S\ \ref{sOBS}\ originate in gas that is quite hot --- if
taken at face value, in gas with temperatures greater than about 1200 K
and 2500 K based on the spectral lines with widths of 0.5 
($\delta$ = 1.8 $\times$ \pow{10}{-4}, and 8.5 $\times$ \pow{10}{-4}) and
0.7 \kmS\ ($\delta$ = 2.4 $\times$ \pow{10}{-4}), respectively.

For simplicity, the foregoing analysis has been based on a two-level maser
transition and the linear maser idealization. The 22 GHz transition
probably is a blend of three hyperfine components. Though this can
influence the analysis in detail, our initial calculations indicate that
the basic conclusion that the observations imply $I_p\ \lesssim$ 0.3 and
high gas temperatures greater than about 1000 K will be unaltered. Perhaps
surprisingly, the line profiles that result from the three merged
hyperfine components still tend to be good Gaussians except at the lowest
temperatures. Again for simplicity, we have not dealt explicitly with the
possibility that the rate $\Gamma_{v}$ for the relaxation of molecular
velocities by the trapping of infrared radiation can be greater than the
decay rate $\Gamma$ of the state (\citealt{gk74}). If so, $\Gamma_{v}$
will replace $\Gamma$ in our calculations (\citealt{nw91}). However,
$\Gamma$ will be reduced by the trapping, and $\Gamma_{v}$ will have
approximately the rate (2 \pow{s}{-1}) that we have used in the foregoing
discussion (\citealt{aw93}). Any changes in our conclusions are expected
to be minimal. The linear maser idealization is considered to be
applicable because astrophysical masers (especially the 22 GHz masers) are
thought to be highly elongated with very small beaming angles so that the
rays are nearly parallel. The $\Delta\Omega\lesssim$
\pow{10}{-4}\ ster inferred above supports this premise for the
masers of \S\ \ref{sOBS}. We cannot, of course, exclude the possibility
that velocity and/or excitation gradients are present within the maser and
have just the right variation to enhance the linewidth while maintaining
a near-Gaussian shape for the overall profile. It seems unlikely that a
velocity gradient which is simply constant along the line of sight will
do this. A velocity gradient will tend to flatten the peak and broaden the
overall profile. When $I_{p} \gtrsim 0.3$ in Figure\ \ref{fKL}, the peak
of the profile already is more flat than a Gaussian. A constant velocity
gradient in this regime would only increase the deviation from a Gaussian.
When $I_{p} \ll 1$ and the maser is unsaturated, a velocity gradient that
is large enough to increase the linewidth significantly would be expected
to cause excessive distortion in the line profle. The remaining
possibility in a constant velocity gradient is that velocity relaxation
is significant  and that $\Gamma_{v}$ is greater than the rate for
stimulated emission, which in turn is greater than $\Gamma$. This would
also correspond to the regime $I_{p} \lesssim 0.3$
in Figure\ \ref{fKL}. Although the peak of the profile does not seem to
be flattened in this case, the available calculations (\citealt{nw88})
also suggest that the velocity gradient causes little increase in the
linewidth --- at least near the minimum for the linewidth in
Figure\ \ref{fKL}. Our interpretation of the data in Figure\ \ref{fMP}\
would then be largely unchanged. Further, detailed calculations as a 
function of the relevant parameters are desirable, however.

\section{DISCUSSION}

An analysis of the line strengths of the masing transitions of water at
higher frequencies by \citet{mmp93}\ has indicated that these masers
originate in a gas that is quite hot ($\gtrsim$ 900 K). Although 22 GHz
\water\ masers do occur along the same lines of sight, the possibility
that they are at different locations along those lines and might originate
in gas at the lower temperatures could not be excluded. Our interpretation
of the Gaussian nature of the 22 GHz profiles indicates that strong 22 GHz
\water\ masers do originate in gas at similarly high temperatures.
Shock waves have long been recognized as likely environments for
interstellar water masers (\citealt*{ssm76}). Recent attention has been
focussed on the specific properties of J-type shocks and C-type shocks. In
the former, the 22 GHz masing has been found to originate in a relatively
narrow range of temperatures near 400 K at which hydrogen molecules
recombine. In contrast, 22 GHz masing at gas temperatures up to 3000 K and
higher can occur in C-type shocks (\citealt{kn96}). Evidently, our analysis
is compatible with an origin for the masers in C-type shocks, but not
J-type shocks. Another implication of the relatively low degree of
saturation of these masers is that the standard Zeeman interpretation of
their observed circular polarization (\citealt{ww01}) is more likely to
be valid (\citealt{str01}). Clearly, more extensive observational efforts
will be necessary to determine whether 22 GHz masers, and other masing
species, commonly have near-Gaussian spectral profiles which can be used
as diagnostic tools.

\acknowledgments

We are grateful for NSF support through Grants AST98-20641 and 
AST99-88104, and to R.~M.~Crutcher, D.~N.~Neufeld, and H.~W.~Wyld for 
helpful comments.


\begin{thebibliography}{}


\bibitem[Anderson \& Watson(1993)]{aw93} Anderson, N., \& Watson, W.~D.\
1993, \apj, 407, 620

\bibitem[Claussen et al.(1994)]{cgjw94} Claussen, M.~J., Gaume, R.~A., 
Johnston, K.~J., \& Wilson, T.~L.\ 1994, \apjl, 424, L41

\bibitem[Emmering \& Watson(1994)]{ew94} Emmering, R.~T., \& Watson,
W.~D.\ 1994, \apj, 424, 991 

\bibitem[Goldreich \& Kwan(1974)]{gk74} Goldreich, P., \& Kwan, J.\ 1974,
\apj, 190, 27

\bibitem[Kaufman \& Neufeld(1996)]{kn96} Kaufman, M.~J., \& Neufeld, D.~A.\
1996, \apj, 456, 250

\bibitem[Litvak(1971)]{lit71} Litvak, M.~M.\ 1971, \apj, 170, 71

\bibitem[Melnick et al.(1993)]{mmp93} Melnick, G.~J., Menten, K.~M.,
Phillips, T.~G., \& Hunter, T.\ 1993, \apjl, 416, L37

\bibitem[Nedoluha \& Watson(1988)]{nw88} Nedoluha, G.~E., \& Watson, W.~D.\
1988, \apjl, 335, L19

\bibitem[Nedoluha \& Watson(1991)]{nw91} Nedoluha, G.~E., \& Watson, W.~D.\
1991, \apjl, 367, L63

\bibitem[Press et al.(1986)Press, Flannery, \& Teukolsky]{pft86} Press,
W.~H., Flannery, B.~P., \& Teukolsky, S.~A.\ 1986, Numerical Recipes:
The Art of Scientific Computing (Cambridge: Cambridge Univ. Press), 457

\bibitem[Sarma et al.(2001)Sarma, Troland, \& Romney]{str01} Sarma, A.~P.,
Troland, T.~H., \& Romney, J.~D.\ 2001, \apjl, 554, L217

\bibitem[Shmeld et al.(1976)Shmeld, Strelnitskii, \& Muzylev]{ssm76} 
Shmeld, I.~K., Strelnitskii, V.~S., \& Muzylev, V.~V.\ 1976, \azh, 53, 728

\bibitem[Tieftrunk et al.(1997)]{tgc97} Tieftrunk, A.~R., Gaume, R.~A.,
Claussen, M.~J., Wilson, T.~L., \& Johnston, K.~J.\ 1997, \aap, 318, 931

\bibitem[Watson \& Wyld(2001)]{ww01} Watson, W.~D., \& Wyld, H.~W.\ 2001,
\apjl, 558, L55

\end{thebibliography}
\end{document}